\documentclass[manyauthors,nocleardouble,COMPASS]{cernphprep} 
\RequirePackage{lineno}
\usepackage{cernunits,cernchemsym,heppennames2,graphicx,subfig,amssymb,color,amsmath}
\usepackage{wrapfig,rotating}
\usepackage{amssymb,amsmath,array,array}
\usepackage{placeins}
\usepackage{cite}

\usepackage[bottom]{footmisc}

\newcommand{\be}{\begin{equation}}
\newcommand{\ee}{\end{equation}}
\newcommand{\bq}{\begin{eqnarray}}
\newcommand{\eq}{\end{eqnarray}}

\begin{document}
\begin{titlepage}
\PHnumber{2019--xxx}
\PHdate{\today}
\DEFCOL{CDS-Library}
\title{Contribution of exclusive  diffractive
processes to the measured  azimuthal asymmetries in SIDIS}
\Collaboration{The COMPASS Collaboration}
\ShortAuthor{The COMPASS Collaboration}

%
%
\section*{The COMPASS Collaboration}
\label{app:collab}
\renewcommand\labelenumi{\textsuperscript{\theenumi}~}
\renewcommand\theenumi{\arabic{enumi}}
\begin{flushleft}
J.~Agarwala\Irefn{triest_i}\Aref{JA},
M.G.~Alexeev\Irefnn{turin_u}{turin_i},
G.D.~Alexeev\Irefn{dubna}, 
A.~Amoroso\Irefnn{turin_u}{turin_i},
V.~Andrieux\Irefnn{cern}{illinois},
N.V.~Anfimov\Irefn{dubna}, 
V.~Anosov\Irefn{dubna}, 
K.~Augsten\Irefnn{dubna}{praguectu}, 
W.~Augustyniak\Irefn{warsaw},
C.D.R.~Azevedo\Irefn{aveiro},
B.~Bade{\l}ek\Irefn{warsawu},
F.~Balestra\Irefnn{turin_u}{turin_i},
M.~Ball\Irefn{bonniskp},
J.~Barth\Irefn{bonniskp},
R.~Beck\Irefn{bonniskp},
Y.~Bedfer\Irefn{saclay},
J.~Berenguer~Antequera\Irefnn{turin_u}{turin_i},
J.~Bernhard\Irefnn{mainz}{cern},
M.~Bodlak\Irefn{praguecu},
P.~Bordalo\Irefn{lisbon}\Aref{A},
F.~Bradamante\Irefn{triest_i},
A.~Bressan\Irefnn{triest_u}{triest_i},
M.~B\"uchele\Irefn{freiburg},
V.~E.~Burtsev\Irefn{tomsk},
W.-C.~Chang\Irefn{taipei},
C.~Chatterjee\Irefnn{triest_u}{triest_i},
M.~Chiosso\Irefnn{turin_u}{turin_i},
A.~G.~Chumakov\Irefn{tomsk},
S.-U.~Chung\Irefn{munichtu}\Aref{B}\Aref{B1}
A.~Cicuttin\Irefn{triest_i}\Aref{C}
P.~M.~M.~Correia\Irefn{aveiro},
M.L.~Crespo\Irefn{triest_i}\Aref{C}
% ...
 \end{flushleft}
%
%
\begin{Authlist}
\item \Idef{aveiro}{University of Aveiro, Dept.\ of Physics, 3810-193 Aveiro, Portugal}
\item \Idef{bochum}{Universit\"at Bochum, Institut f\"ur Experimentalphysik, 44780 Bochum, Germany\Arefs{Q}$^,$\Arefs{R}}
\item \Idef{bonniskp}{Universit\"at Bonn, Helmholtz-Institut f\"ur  Strahlen- und Kernphysik, 53115 Bonn, Germany\Arefs{Q}}
\item \Idef{bonnpi}{Universit\"at Bonn, Physikalisches Institut, 53115 Bonn, Germany\Arefs{Q}}
\item \Idef{brno}{Institute of Scientific Instruments, AS CR, 61264 Brno, Czech Republic\Arefs{S}}
\item \Idef{calcutta}{Matrivani Institute of Experimental Research \& Education, Calcutta-700 030, India\Arefs{T}}
\item \Idef{dubna}{Joint Institute for Nuclear Research, 141980 Dubna, Moscow region, Russia\Arefs{E}} 
\item \Idef{freiburg}{Universit\"at Freiburg, Physikalisches Institut, 79104 Freiburg, Germany\Arefs{Q}$^,$\Arefs{R}}
\item \Idef{cern}{CERN, 1211 Geneva 23, Switzerland}
\item \Idef{liberec}{Technical University in Liberec, 46117 Liberec, Czech Republic\Arefs{S}}
\item \Idef{lisbon}{LIP, 1649-003 Lisbon, Portugal\Arefs{U}}
\item \Idef{mainz}{Universit\"at Mainz, Institut f\"ur Kernphysik, 55099 Mainz, Germany\Arefs{Q}}
\item \Idef{miyazaki}{University of Miyazaki, Miyazaki 889-2192, Japan\Arefs{V}}
\item \Idef{moscowlpi}{Lebedev Physical Institute, 119991 Moscow, Russia}
\item \Idef{munichtu}{Technische Universit\"at M\"unchen, Physik Dept., 85748 Garching, Germany\Arefs{Q}$^,$\Arefs{D}}
\item \Idef{nagoya}{Nagoya University, 464 Nagoya, Japan\Arefs{V}}
\item \Idef{praguecu}{Charles University in Prague, Faculty of Mathematics and Physics, 18000 Prague, Czech Republic\Arefs{S}}
\item \Idef{praguectu}{Czech Technical University in Prague, 16636 Prague, Czech Republic\Arefs{S}}
\item \Idef{protvino}{State Scientific Center Institute for High Energy Physics of National Research Center `Kurchatov Institute', 142281 Protvino, Russia}
\item \Idef{saclay}{IRFU, CEA, Universit\'e Paris-Saclay, 91191 Gif-sur-Yvette, France\Arefs{R}}
\item \Idef{taipei}{Academia Sinica, Institute of Physics, Taipei 11529, Taiwan\Arefs{W}}
\item \Idef{telaviv}{Tel Aviv University, School of Physics and Astronomy, 69978 Tel Aviv, Israel\Arefs{X}}
\item \Idef{triest_u}{University of Trieste, Dept.\ of Physics, 34127 Trieste, Italy}
\item \Idef{triest_i}{Trieste Section of INFN, 34127 Trieste, Italy}
\item \Idef{turin_u}{University of Turin, Dept.\ of Physics, 10125 Turin, Italy}
\item \Idef{turin_i}{Torino Section of INFN, 10125 Turin, Italy}
\item \Idef{tomsk}{Tomsk Polytechnic University, 634050 Tomsk, Russia\Arefs{Y}}
\item \Idef{illinois}{University of Illinois at Urbana-Champaign, Dept.\ of Physics, Urbana, IL 61801-3080, USA\Arefs{Z}}
\item \Idef{warsaw}{National Centre for Nuclear Research, 02-093 Warsaw, Poland\Arefs{a} }
\item \Idef{warsawu}{University of Warsaw, Faculty of Physics, 02-093 Warsaw, Poland\Arefs{a} }
\item \Idef{warsawtu}{Warsaw University of Technology, Institute of Radioelectronics, 00-665 Warsaw, Poland\Arefs{a} }
\item \Idef{yamagata}{Yamagata University, Yamagata 992-8510, Japan\Arefs{V} }
\item \Idef{retired}{Retired}
\end{Authlist}
%
%
\renewcommand\theenumi{\alph{enumi}}
\begin{Authlist}
\item [{\makebox[2mm][l]{\textsuperscript{\#}}}] Corresponding authors
\item \Adef{JA}{Present address: University of Pavia, 27100 Pavia, Italy}
\item \Adef{A}{Also at Instituto Superior T\'ecnico, Universidade de Lisboa, Lisbon, Portugal}
\item \Adef{B}{Also at Dept.\ of Physics, Pusan National University, Busan 609-735, Republic of Korea}
\item \Adef{B1}{Also at at Physics Dept., Brookhaven National Laboratory, Upton, NY 11973, USA}
\item \Adef{C}{Also at Abdus Salam ICTP, 34151 Trieste, Italy}
\item \Adef{D}{Supported by the DFG cluster of excellence `Origin and Structure of the Universe' (www.universe-cluster.de) (Germany)}
\item \Adef{F}{Supported by the DFG Research Training Group Programmes 1102 and 2044 (Germany)}
\item \Adef{G}{Also at Chubu University, Kasugai, Aichi 487-8501, Japan\Arefs{V}}
\item \Adef{H}{Also at Dept.\ of Physics, National Central University, 300 Jhongda Road, Jhongli 32001, Taiwan}
\item \Adef{I}{Also at KEK, 1-1 Oho, Tsukuba, Ibaraki 305-0801, Japan}
\item \Adef{J}{Present address: Universit\"at Bonn, Physikalisches Institut, 53115 Bonn, Germany}
\item \Adef{K}{Also at Moscow Institute of Physics and Technology, Moscow Region, 141700, Russia}
\item \Adef{L}{Also at Yerevan Physics Institute, Alikhanian Br. Street, Yerevan, Armenia, 0036}
\item \Adef{M}{Also at Dept.\ of Physics, National Kaohsiung Normal University, Kaohsiung County 824, Taiwan}
\item \Adef{b}{Supported by ARN,  France with the P2IO LabEx (ANR-10-LBX-0038) in the framework ``Investissements d'Avenir'' (ANR-11-IDeX-003-01)}
\item \Adef{N}{Also at Institut f\"ur Theoretische Physik, Universit\"at T\"ubingen, 72076 T\"ubingen, Germany}
\item \Adef{O}{Also at University of Eastern Piedmont, 15100 Alessandria, Italy}
\item \Adef{P}{Present address: RWTH Aachen University, III.\ Physikalisches Institut, 52056 Aachen, Germany}
%
%
\item \Adef{Q}{Supported by BMBF - Bundesministerium f\"ur Bildung und Forschung (Germany)}
\item \Adef{R}{Supported by FP7, HadronPhysics3, Grant 283286 (European Union)}
\item \Adef{S}{Supported by MEYS, Grant LM20150581 (Czech Republic)}
\item \Adef{T}{Supported by B.Sen fund (India)}
\item \Adef{E}{Supported by CERN-RFBR Grant 12-02-91500}
\item \Adef{U}{Supported by FCT, Grants CERN/FIS-PAR/0007/2017 and  CERN/FIS-PAR/0022/2019 (Portugal)}
\item \Adef{V}{Supported by MEXT and JSPS, Grants 18002006, 20540299, 18540281 and 26247032, the Daiko and Yamada Foundations (Japan)}
\item \Adef{W}{Supported by the Ministry of Science and Technology (Taiwan)}
\item \Adef{X}{Supported by the Israel Academy of Sciences and Humanities (Israel)}
\item \Adef{Y}{Supported by the Russian Federation  program ``Nauka'' (Contract No. 0.1764.GZB.2017) (Russia)}
\item \Adef{Z}{Supported by the National Science Foundation, Grant no. PHY-1506416 (USA)}
\item \Adef{a}{Supported by NCN, Grant 2017/26/M/ST2/00498 (Poland)}
\end{Authlist}

%



\begin{abstract}
Hadron leptoproduction in  Semi-Inclusive measurements of
Deep-Inelastic 
Scattering (SIDIS) on unpolarised nucleons allows one to get information on the
intrinsic transverse momentum of quarks in a 
nucleon and on the Boer-Mulders function through the measurement of  
azimuthal modulations in the cross section. 
These modulations
were recently measured by the HERMES experiment at DESY on proton
and deuteron targets,
and by the COMPASS experiment using the CERN SPS muon beam and a $^6$LiD
target.
In both cases, the amplitudes of the $\cos\phi_h$ and $\cos 2\phi_h$
modulations  show strong kinematic dependences
for both positive and negative hadrons. 
It has been known since some time that the measured final-state
hadrons in those SIDIS experiments receive a 
contribution from exclusive diffractive production of 
vector mesons, particularly 
important at large values of 
$z$, the fraction of the virtual photon energy carried by the hadron. 
In previous measurements of azimuthal asymmetries 
this contribution was not taken into account, because it was
not known that it could distort the azimuthal modulations. 
Presently, a method to evaluate the contribution of 
the exclusive reactions  to the
azimuthal asymmetries measured by COMPASS has been developed.
The subtraction of this contribution
results in a better understanding of the kinematic effects,
and the remaining non-zero $\cos 2\phi_h$ modulation 
gives indication for a non-zero Boer-Mulders effect.
\end{abstract}

\vfill
\Submitted{(to be submitted as a Rapid Communication to Nuclear Physics B)}
\end{titlepage}

\section{Introduction}

The azimuthal asymmetries in Semi-Inclusive measurements of
Deep-Inelastic Scattering (SIDIS) on unpolarised nucleons
are a powerful tool to access the quark intrinsic transverse momentum 
$k_T$ and
the Boer-Mulders~\cite{Boer:1997nt} Transverse Momentum Dependent Parton 
Distribution Function (TMD PDF)  $h_1^{\perp}$. 
The target spin-averaged differential SIDIS cross section for the production 
of a
hadron $h$ is given in the one-photon exchange
approximation~\cite{bacchetta} by~\footnote{
In this paper we use the same notation as in Ref.~\cite{Adolph:2014pwc}.
}
\begin{eqnarray}
\frac{\textrm{d}\sigma}{p_T^{\,h}\,\textrm{d}p_T^{\,h}\,\textrm{d}x\,\textrm{d}y\, \textrm{d}z\, \textrm{d}\phi_h} &=&
\sigma_0 \, \Big( 1 + \epsilon _1 A^{UU}_{\cos\phi_h} \cos \phi_h + 
\nonumber \\
&& + \epsilon _2 A^{UU}_{\cos 2\phi_h} \cos2\phi_h + \lambda \epsilon _3 A^{LU}_{\sin\phi_h} \sin\phi_h  \Big) \;,
\label{eq:cross_section_1}
\end{eqnarray}
where $\phi_h$ is the azimuthal angle of the hadron with respect to the lepton
scattering plane,
in a reference system in which the z-axis is the virtual photon
direction and the x-axis is defined by the scattered lepton transverse 
momentum.
The transverse momentum $p_T^{\,h}$ of the hadron is the component of
$\vec{p}^{\,h}$ orthogonal to the z-axis and $z$ is the fraction of 
the available energy carried by the hadron.
The quantity $x$ is the Bjorken variable, $y$ is  the
fractional energy of the virtual photon,
$\sigma_0$ is the $\phi_h$-independent part of the cross section,
$\lambda$ is the longitudinal polarisation of the incident lepton,
and $\epsilon_1, \, \epsilon_2$ and $\epsilon_3$ are kinematic 
factors depending on $y$.
The amplitudes $A^{XU}_{f(\phi_h)}$ are referred to as azimuthal asymmetries
in the following. 
The superscripts \emph{UU} and \emph{LU} refer to unpolarised
beam and target, and to longitudinally polarised beam and unpolarised target,
respectively.
In particular, within the pQCD factorized approach~\cite{bacchetta}, 
the twist-3 azimuthal
asymmetry $A^{UU}_{\cos\phi_h}$ gives a direct access to $\langle k_T^2 \rangle$ 
through the Cahn effect~\cite{Cahn:1978se},
which is expected to be the main contributor to $A^{UU}_{\cos\phi_h}$.
The twist-2 part of
the asymmetry $A^{UU}_{\cos 2\phi_h}$ gives access to the Boer-Mulders TMD PDF.

Measurements of the ``unpolarised'' SIDIS azimuthal asymmetries were
recently performed by 
the HERMES Collaboration for charged
hadrons, pions and kaons using both proton and deuteron 
targets~\cite{Airapetian:2012yg},
and by the COMPASS Collaboration  for charged hadrons
using a deuteron ($^6$LiD) target~\cite{Adolph:2014pwc}. 
They all show strong dependences on the kinematic variables.
Several phenomenological analyses (for more details see 
Ref.~\cite{Barone:2015ksa}) did not succeed either in reproducing the data or in extracting the Boer-Mulders PDF. 
As a result the present knowledge of the quark intrinsic transverse 
momentum has very large uncertainties and a possible non-zero 
Boer-Mulders function in the SIDIS cross section has still to be 
demonstrated.

Looking at the COMPASS results, a few aspects for the 
$A^{UU}_{\cos \phi_h}$ asymmetry are
particularly intriguing.
This asymmetry is expected to be mainly due to the kinematic
Cahn effect and should be negative, with absolute value 
increasing
almost linearly with $z$ and $p_T^h$ 
and proportional  to $\langle k_T^2 \rangle$
(see e.g. Ref.~\cite{Barone:2015ksa}).
The trend of the data is, however, quite different.
The measured $z$ dependence of the integrated 
asymmetry~\footnote{
See fig. 10 of Ref.~\cite{Adolph:2014pwc}}
shows a strong increase of absolute value starting at $z \simeq 0.5$.
Moreover, looking at the three-dimensional 
result~\footnote{
See fig. 12 of Ref.~\cite{Adolph:2014pwc}}, at high $z$
the $p_T^h$ dependence is the opposite of the expected one, and 
the $x$ dependence changes behaviour from low to high $z$.

These observations suggest that another mechanism, different from
the TMD parton model, is at work in  hadron production
at large $z$. 
As a matter of fact
it is known that the charged hadron SIDIS sample at large $z$ and at small 
$p_T^h$ contains a non-negligible contribution of hadrons from the decay 
of vector mesons (VM) produced in exclusive diffractive processes.
This contribution was  indeed  taken into account in the
measurements of hadron 
multiplicities~\cite{Airapetian:2012ki,Adolph:2016bga,Adolph:2016bwc,Aghasyan:2017ctw,Akhunzyanov:2018ysf}.
Now, for the first time,
we have  investigated the effect of  this VM contribution on the azimuthal
asymmetries.
We have measured azimuthal asymmetries 
for $h^+$ and $h^-$ originating from the decay of exclusively produced VMs 
(referred to in the following as "exclusive-VM hadrons''),
and found them to be large.
Since they do not have an interpretation in the framework of the parton model
TMD formalism, we have subtracted this contribution from the published 
COMPASS asymmetries. 
This correction considerably improves the agreement with the expectations
for $A^{UU}_{\cos \phi_h}$ and has also a noticeable effect for
$A^{UU}_{\cos 2\phi_h}$.

The paper is organized as follows:
in Section 2 the measurement of the azimuthal modulations for exclusive-VM 
hadrons is described.
In Section 3 we present the calculation of the fraction of exclusive-VM hadrons 
in the measured hadron sample.
In Section 4 we describe the procedure used to subtract the exclusive-VM hadron 
contribution to the
azimuthal asymmetries published by COMPASS, and give the final 
results.

\section{Azimuthal modulations of exclusive-VM hadron }

In order to evaluate the contribution of exclusive-VM hadrons to the published
azimuthal asymmetries~\cite{Adolph:2014pwc} obtained from the COMPASS data
collected in 2004, we have analysed the 2006 COMPASS data, which were recently 
used to measure the hadron multiplicities in 
SIDIS~\cite{Adolph:2016bga,Adolph:2016bwc,Aghasyan:2017ctw,Akhunzyanov:2018ysf}, 
and for which
all the necessary simulated data are available.
The experimental conditions of the two data sets are very similar,
since the same target material ($^6$LiD) was used, once limiting the
spectrometer acceptance to the same restricted kinematic region
investigated in Ref.~\cite{Adolph:2014pwc}.

\begin{figure}[t]
\includegraphics[width=0.45\textwidth]{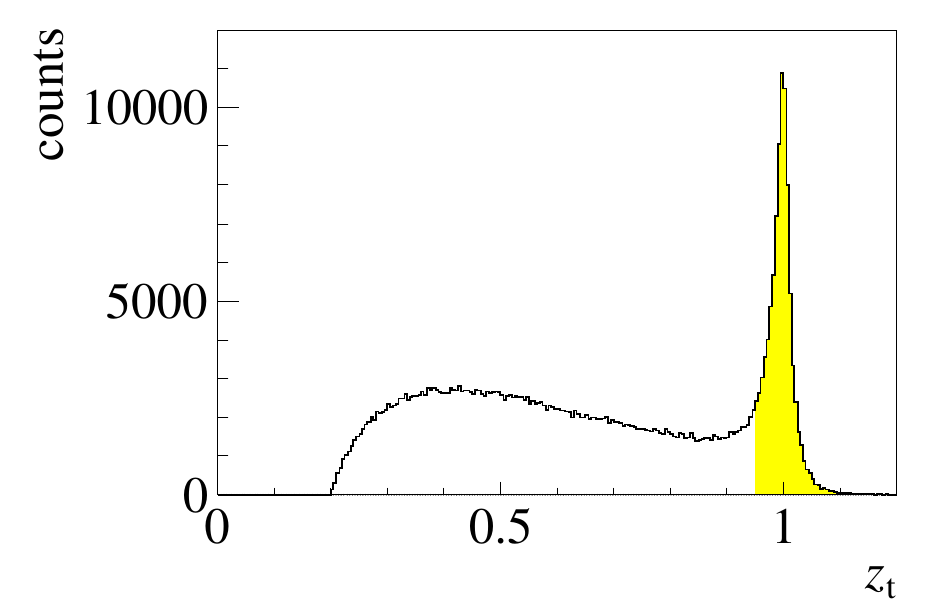}
\hfill
\includegraphics[width=0.45\textwidth]{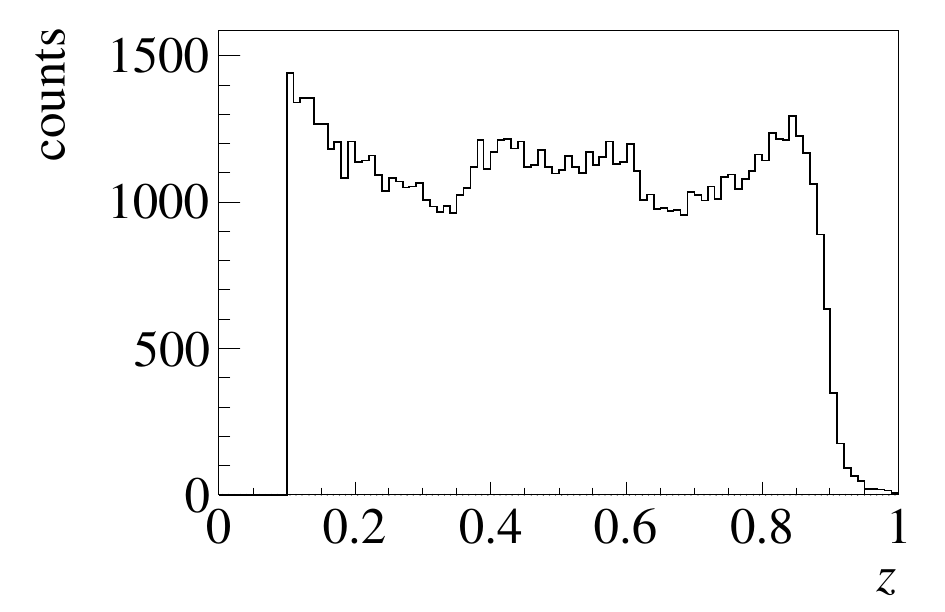}
\caption{Left panel: distribution of $z_t$ for the events with only two reconstructed hadrons with opposite charge. 
The exclusive events are selected by the cut $z_t>0.95$.
Right panel: $z$ distribution for the positive hadron of the selected pairs.}\label{fig:zt}
\end{figure}

The azimuthal modulations of the exclusive-VM hadrons 
are measured selecting DIS events as
in Ref. \cite{Adolph:2014pwc}, i.e. by using:
\begin{center}
$Q^2>1$ (GeV/c)$^2$, $\, W>5$ GeV/c$^2$, $\, 0.2<y<0.9$, 
\end{center}
where $Q^2$ is the exchanged photon virtuality and $W$ the final 
state hadronic 
mass.
The  events are then selected requiring 
in the final state, in addition to the scattered muon,
only two oppositely 
charged hadrons with $z>0.1$.
The fraction of the final-state energy that is carried by the hadron pair,
 $z_t$, is shown in the left panel of 
Fig. \ref{fig:zt}. 
Hadron pairs originating from exclusively produced vector mesons 
appear as the sharp peak at $z_t \simeq 1$
and  are selected by requiring $z_t > 0.95$.
Contributions from other processes, which appear as background to this peak, 
are neglected in the present analysis.

The $z$ distribution for the positive hadrons of the selected
pairs is shown in the right panel of the same figure. 
Most of the hadrons come from $\rho^0$ decays.
The broad structure at $0.4 < z < 0.6$ is due to 
hadrons from $\phi$ 
meson decays, whose contribution is less than 10\% of that of the $\rho^0$.

\begin{figure}[tb]
\includegraphics[width=0.45\textwidth]{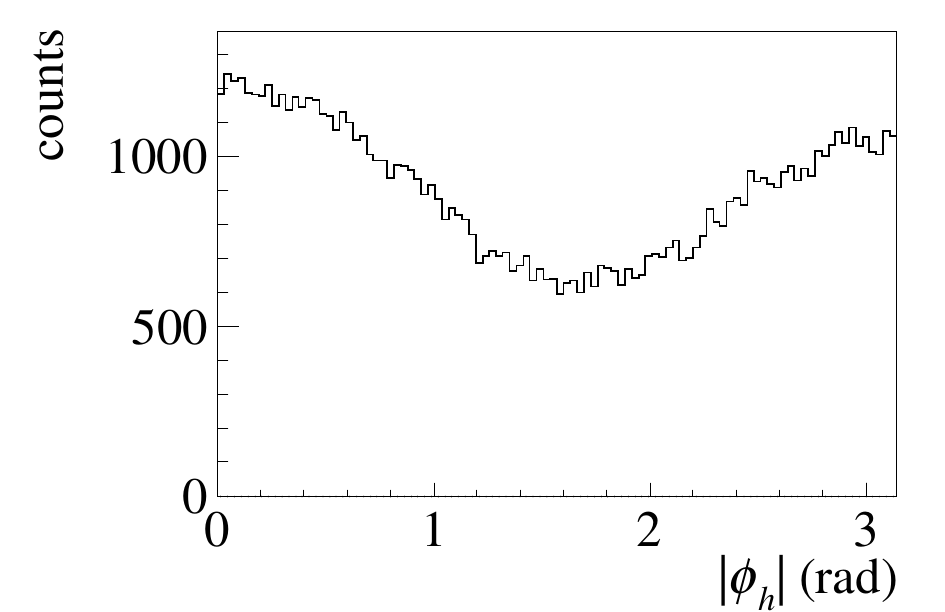}
\hfill
\includegraphics[width=0.45\textwidth]{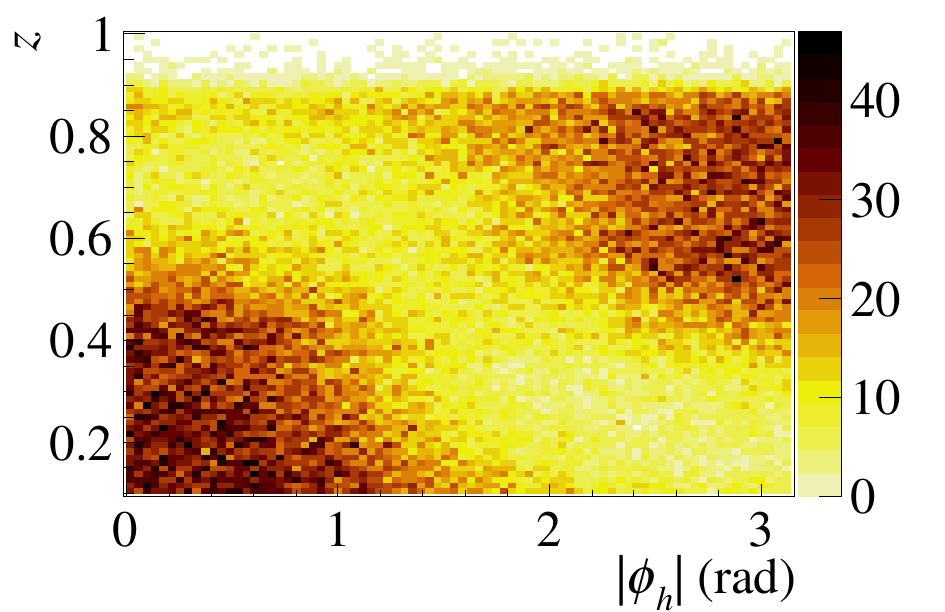}
\caption{Distribution of $|\phi_h|$ (left panel) and correlation between $z$ and $|\phi_h|$ (right panel) for positive exclusive-VM hadrons.}
\label{fig:phih}
\end{figure}

The $|\phi_h|$ distribution of the exclusive-VM hadrons shows
large modulations, as can be seen 
in the left panel of Fig.~\ref{fig:phih} for positive hadrons. 
Furthermore the $|\phi_h|$ distribution strongly depends on $z$, 
as can be seen from the right panel in Fig.~\ref{fig:phih}, again for $h^+$.
From that 2-dimensional distribution 
one notices that the amplitude of the $\cos\phi_h$ modulation changes sign with $z$. 
The same properties are observed also for $h^-$.

The acceptance-corrected azimuthal modulations of the positive and negative 
exclusive-VM hadrons are fitted in each $x$, $z$ and $p_T^h$ 
bin of Ref.~\cite{Adolph:2014pwc} with the function 
\begin{equation}\label{eq:fexcl}
    f(\phi_h)=a_0[1+\epsilon_1 a_1\cos\phi_h +\epsilon_2 a_2\cos 2\phi_h] \, ,
\end{equation}
where the amplitudes $a_0$, $a_1$ and $a_2$ are free parameters.
The $\sin\phi_h$ modulation is not  included because parallel studies 
on exclusive vector-meson production in COMPASS do not exhibit such a 
modulation~\cite{Adolph:2013zaa}. 
Other possible orthogonal modulations are not relevant since 
they do not appear in the SIDIS cross section.

\begin{figure}[tb]
	\centering
	\includegraphics[width=0.99\textwidth]{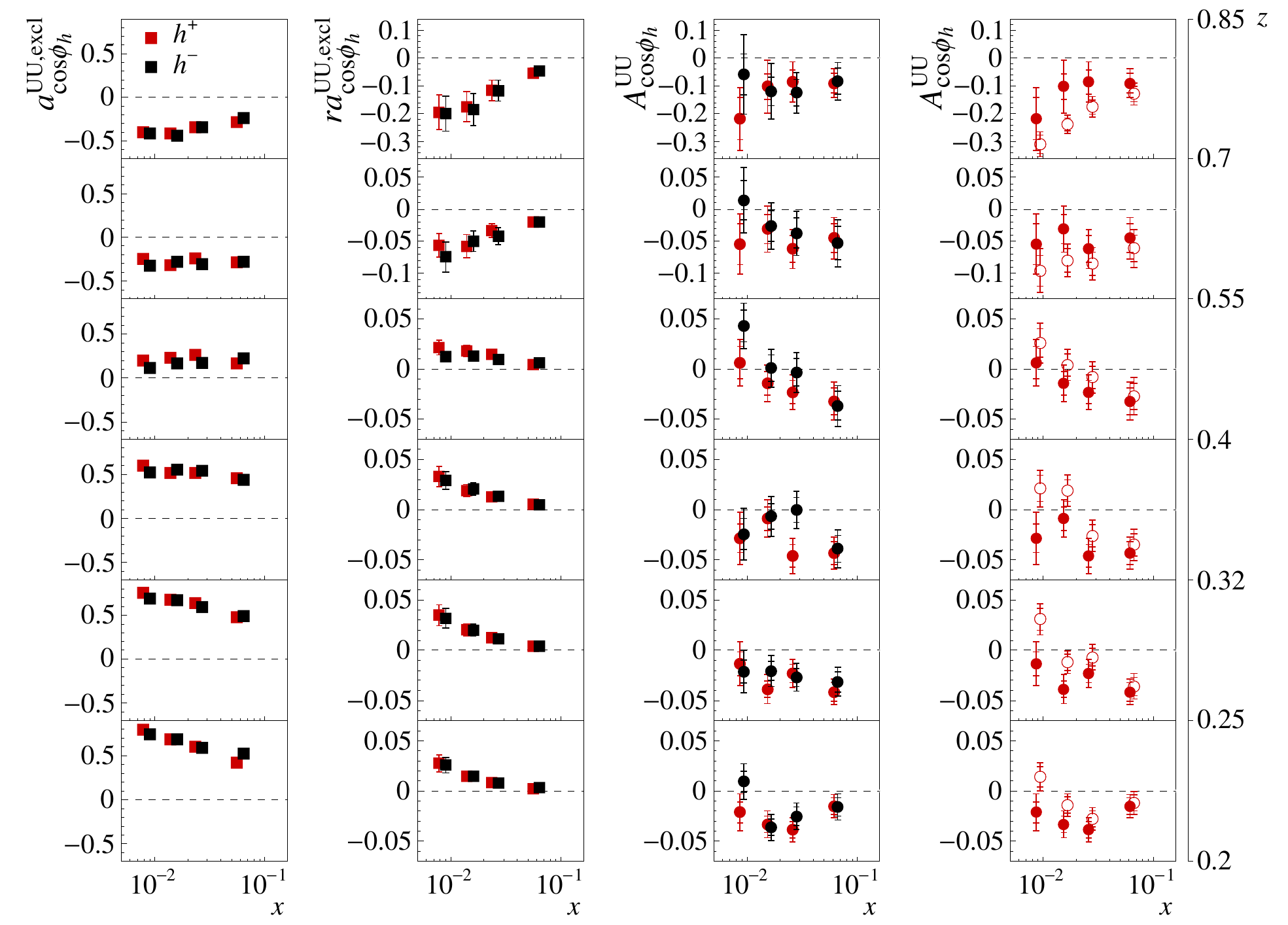}
\caption{First column: $a^{UU,excl}_{\cos\phi_h}$ amplitude for $h^+$ 
(red squares) and $h^-$ (black squares). 
Second column: $r\,a^{UU,excl}_{\cos\phi_h}$ for $h^+$ (red squares) 
and $h^-$ (black squares). 
Third column: $A^{UU}_{\cos\phi_h}$ asymmetry after the subtraction 
of exclusive-VM hadron contribution for $h^+$ (red circles) and $h^-$ (black
circles). Last column: comparison between the 
asymmetry for $h^+$
before (open circles) and after (full circles) exclusive-VM hadron
subtraction. 
From bottom to top, results for increasing values of $z$ are shown, as indicated on the very right of the figure.
All the results refer to the first
$p_T^h$ bin (0.1~GeV/c~$<p_T^h<0.3$~GeV/c).
}
\label{fig:cosphi_all}
\end{figure}

The fitted amplitudes of the $\cos\phi_h$ and $\cos2\phi_h$ modulations for 
exclusive-VM hadrons,
$a^{UU,excl}_{\cos\phi_h}$ and $a^{UU,excl}_{\cos2\phi_h}$,
decrease with increasing $p_T^h$ and are almost equal for $h^+$ and $h^-$, 
indicating that what is modulated is the direction of the parent VM.
As an example, the amplitudes $a^{UU,excl}_{\cos\phi_h}$ 
for 0.1~GeV/c$\, <p_T^h< \,$0.3~GeV/c
are shown in the first column of Fig.~\ref{fig:cosphi_all} 
for both $h^+$ and $h^-$.
The $a^{UU,excl}_{\cos\phi_h}$ amplitude is very large in absolute value
at large and small $z$, and changes sign at $z\simeq 0.5$. 
The $a^{UU,excl}_{\cos 2\phi_h}$ amplitudes are smaller but still non-negligible.  

It should be noted that the 
results of the present analysis refer to a $^6$LiD target and COMPASS 
kinematics.
The observed azimuthal asymmetries for 
exclusive-VM hadrons depend on the angular distributions for $\rho^0$ decay and 
production, which are determined by Spin Density Matrix Elements (SDMEs).
The SDMEs depend on $\rho^0$ transverse momentum ~\cite{Airapetian:2009af} 
and on the mechanism of its production. 
In particular, for coherent production on the target nuclei, which 
dominates at small $p_T^h$, one may expect different angular distributions 
(different SDMEs) than those for the production on a single free or 
quasi-free nucleon.

\section{Fraction of exclusive-VM hadrons in the SIDIS sample}

For a quantitative estimate of the exclusive-VM hadron contribution to the 
unpolarised azimuthal asymmetries, 
it is necessary to determine the number $N_h^{excl}$  of exclusive-VM hadrons 
relative to the total number of hadrons $N_h^{tot}$ , i.e. the ratio
$r=N_h^{excl}/N_h^{tot}$. 
Here we use a parameterisation obtained from previous works 
\cite{Adolph:2016bga,Adolph:2016bwc,Aghasyan:2017ctw}, which was based on 
a combined use of HEPGEN~\cite{Sandacz:2012at} 
and LEPTO~\cite{Ingelman:1996mq} Monte Carlo generators. 
The former one is used to model differential cross sections 
of various hard processes of exclusive leptoproduction of single mesons or
photons at COMPASS kinematics.
For the determination of $r$, only exclusive $\rho^0$  production, which gives
the main contribution to the exclusive-VM hadrons,
is taken into account in the present study.
By doing this, we might underestimate $r$, but only in the bins at lowest $p_T^h$ 
and $z\simeq 0.5$, where it could be larger by at most a factor 1.2.

\begin{figure}[tb]
	\centering
	\includegraphics[width=0.99\textwidth]{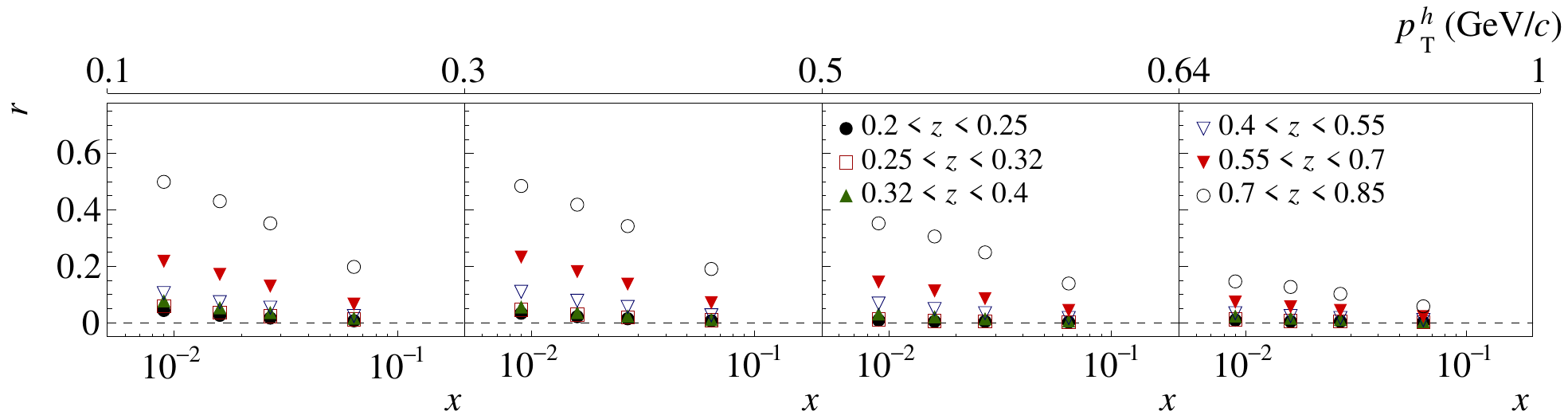}
\caption{Fraction $r$ of exclusive-VM hadrons evaluated as function of
$x$
in the different $z$ and $p_T^h$ bins.}
\label{fig:r}
\end{figure}

Since the binning in 
Ref.~\cite{Adolph:2016bga,Adolph:2016bwc,Aghasyan:2017ctw}
is different from that in Ref.~\cite{Adolph:2014pwc},
we had to parameterise $r$ as a function of
$x$, $z$ and $p_T^h$.
The estimated values of $r$ in all the kinematic bins are shown 
in Fig.~\ref{fig:r} and are assumed to be
the same  for positive and negative 
hadrons.
As one can see, the fraction of pions coming from the decay of exclusively produced $\rho^0$ is very large at large $z$ and small $p_T^h$, where it reaches 50\%, and diminishes for decreasing $z$ and increasing $p_T^h$.
The overall systematic uncertainty on $r$ is estimated to be 
approximately  30\% and is mainly 
due to the uncertainty on the knowledge of the  diffractive 
cross section \cite{Adolph:2016bga,Adolph:2016bwc,Aghasyan:2017ctw}.

\section{Results for the unpolarised SIDIS azimuthal 
asymmetries}
\label{sec:pureSIDIS}

The exclusive-VM hadron contributions to the published azimuthal asymmetries 
$r\,a^{UU,excl}_{\cos \phi_h}$ and $r\,a^{UU,excl}_{\cos 2\phi_h}$ 
are calculated in each
$x$, $z$ and $p_T^h$ bin of Ref.~\cite{Adolph:2014pwc}.
The results for the smallest $p_T^h$ bin, i.e. 
0.1~GeV/c~$< p_T^h<0.3$~GeV/c, are shown for  
$h^+$ and $h^-$ in the second column of Fig.~\ref{fig:cosphi_all}.
As can be seen, the contribution of exclusive-VM hadrons is clearly different 
from zero and reaches values up to 20\% at large $z$ in this low $p_T^h$
range.
The contribution to the $\cos 2\phi_h$ modulation is smaller but still non-negligible, in particular if compared to the measured values of the asymmetries.

The  asymmetries  $A^{UU}_{\cos\phi_h}$, corrected for the contribution
of exclusive-VM hadrons, are obtained using
\begin{eqnarray}\label{eq:pure_A}
    A_{\cos \phi_h}^{UU}&=& \frac{1}{1-r}\left(A_{\cos \phi_h}^{UU, \,
        publ}-r\,a_{\cos \phi_h}^{UU,excl} \right),
\end{eqnarray}
where $A_{\cos i\phi_h}^{UU,\,publ}$ are the published values.
A similar expression is used to obtain $A^{UU}_{\cos2\phi_h}$.

The resulting $A^{UU}_{\cos\phi_h}$  azimuthal asymmetries 
are shown in the third column of Fig.~\ref{fig:cosphi_all}, 
again for the smallest $p_T^h$ bin. 
After subtraction, the $x$ dependence of the asymmetry becomes weaker,
and in particular only a few 
positive values that are hard to be described by the Cahn effect remain.
The last column of the figure shows the comparison between the 
asymmetries as published and after subtracting the 
contribution of exclusive VMs for $h^+$. 
One can also see that the contribution of exclusive-VM hadrons is sizable 
at all $z$. 

\begin{figure}[tb]
	\centering
	\includegraphics[width=0.99\textwidth]{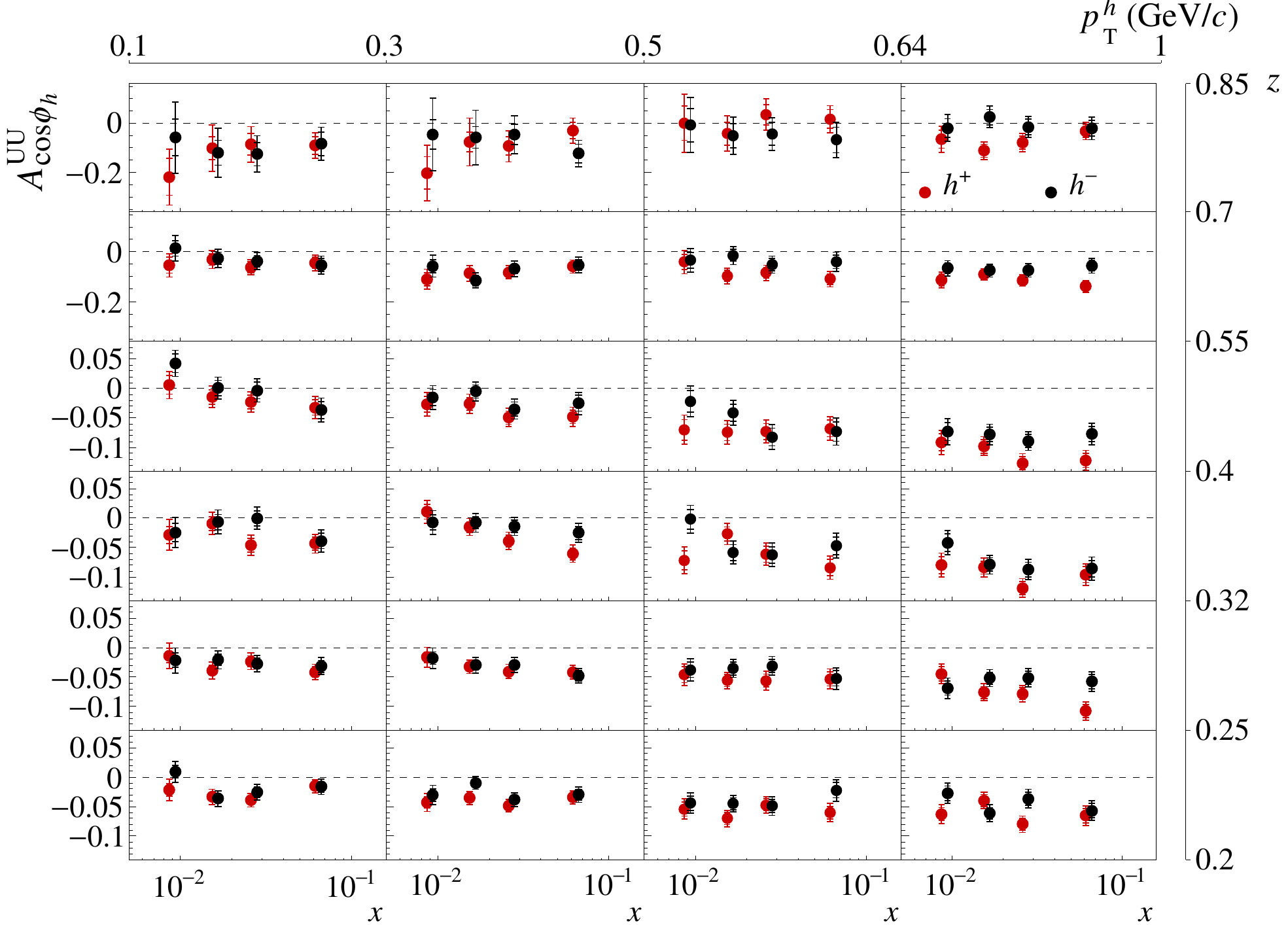}
\caption{SIDIS $A^{UU}_{\cos\phi_h}$ asymmetry on $^6$LiD for $h^+$ 
(red circles) and $h^-$ (black circles)
after subtracting from the published asymmetry \cite{Adolph:2014pwc} 
the contribution of exclusive-VM hadrons,
as function of  $x$, in $z$ and $p_T^h$ bins.
Inner error bars denote statistical uncertainties, outer ones statistical 
and systematic uncertainties added in quadrature.}
\label{fig:sub_cosphi}
\end{figure}

\begin{figure}[tb]
	\centering
	\includegraphics[width=0.99\textwidth]{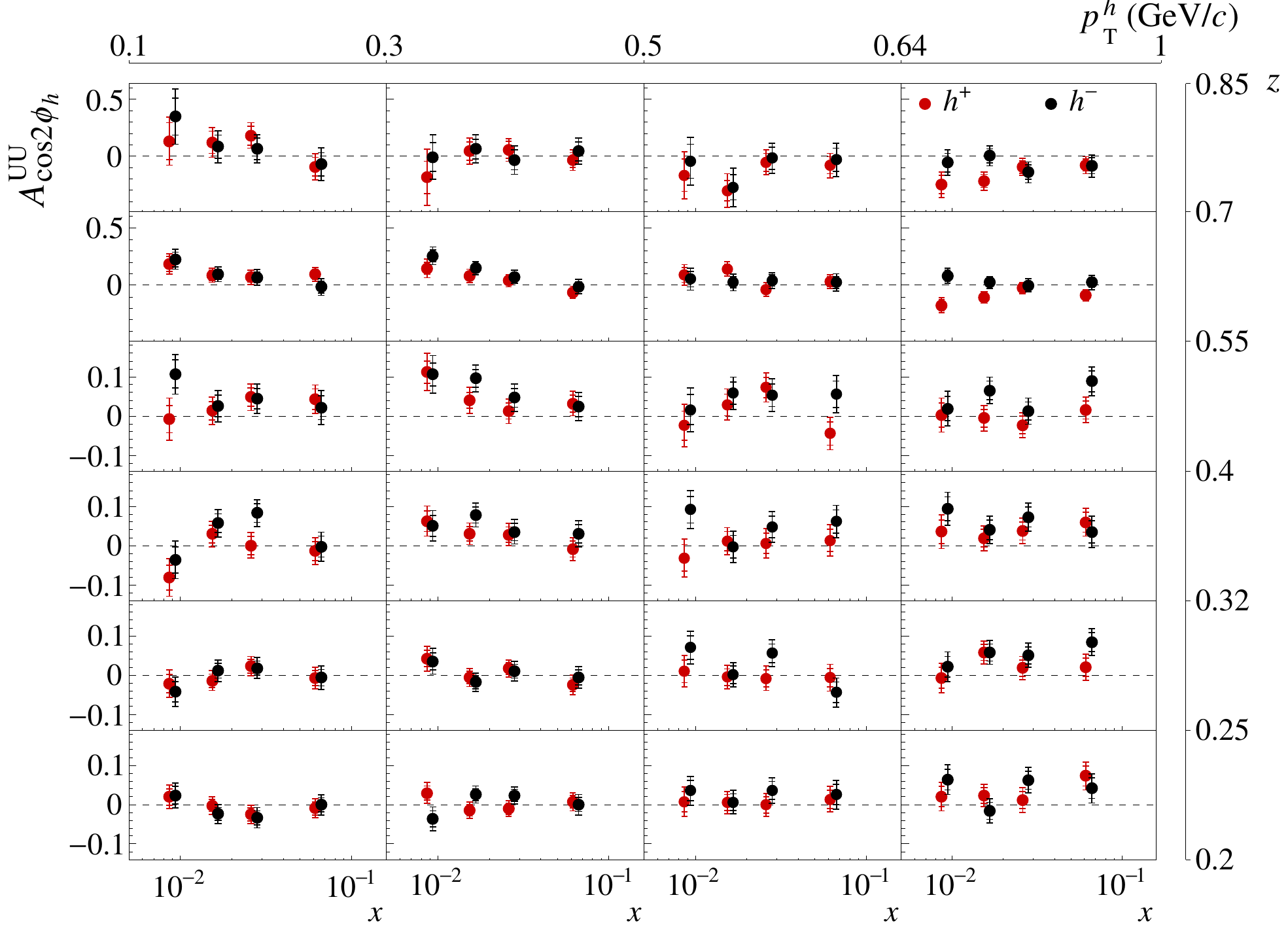}
\caption{SIDIS $A^{UU}_{\cos 2\phi_h}$  asymmetry on $^6$LiD for $h^+$
(red circles) and $h^-$ (black circles)
after subtracting from the published asymmetry \cite{Adolph:2014pwc} 
the contribution of exclusive-VM hadrons,
as function of  $x$, in $z$ and $p_T^h$ bins.
Inner error bars denote statistical uncertainties, outer ones statistical 
and systematic uncertainties added in quadrature.}
\label{fig:sub_cos2phi}
\end{figure}

The  results for 
$A^{UU}_{\cos\phi_h}$ and $A^{UU}_{\cos2\phi_h}$ for positive and negative
hadrons are shown in all  $x$, $z$ and $p_T^h$ bins in 
Fig. \ref{fig:sub_cosphi} and \ref{fig:sub_cos2phi}, respectively.
The inner error bars correspond to the statistical uncertainties only,
while the outer bars represent the total uncertainties.
The increase in the statistical uncertainties is due to the
low statistics of the exclusive-VM hadrons.
The systematic uncertainties have been evaluated by adding 
in quadrature the uncertainties  of the 
published results (estimated to be of the same order of the statistical
ones) and those due to the subtraction procedure.
For the last ones the dominant contribution is that of the
poor knowledge of $r$, which can cause an uncertainty 
at most as large as the statistical one,
apart from a few bins at the highest $z$- and lowest $x$-values.
The total uncertainties are evaluated by adding in quadrature 
the statistical and the systematic uncertainties.
The  numerical values of the asymmetries are available on
HepData~\cite{hepdata}.

\begin{figure}[tb]
	\centering
	\includegraphics[width=0.99\textwidth]{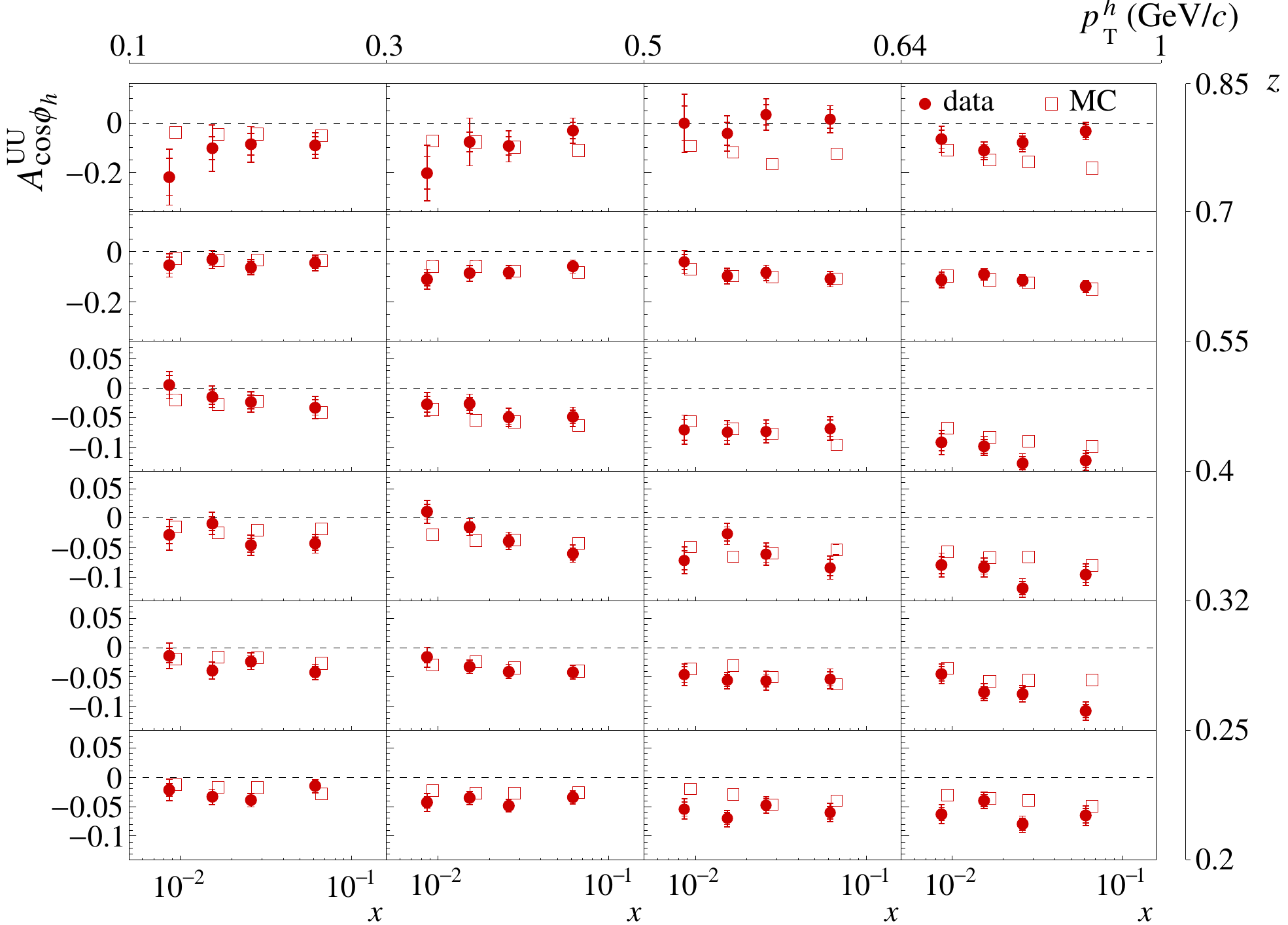}
\caption{Comparison between the SIDIS $A^{UU}_{\cos \phi_h}$ asymmetry,
as function of  $x$, in $z$ and $p_T^h$ bins, for $h^+$
on $^6$LiD after 
 subtracting the exclusive-VM hadron contribution (closed circles) and
the results of a 
Monte Carlo simulation (open squares) which includes the Cahn effect.
Inner error bars denote statistical uncertainties, outer ones statistical and systematic uncertainties added in quadrature.}
\label{fig:sub_MC_cosphi_pos}
\end{figure}

In spite of the large uncertainties we consider this work as a major
step forward in understanding the 3D structure of the nucleon.
To give an idea of the impact, in 
Fig.~\ref{fig:sub_MC_cosphi_pos} we compare $A^{UU}_{\cos\phi_h}$
with a simple Monte Carlo simulation for the Cahn effect.
We have used the Monte Carlo code of Refs.~\cite{Kerbizi:2018qpp,AlbiDubna},
describing the fragmentation of 
polarised quarks, which was modified to include 
the Cahn effect.
This is achieved 
by modulating the fragmenting quark direction according to the 
lepton-quark hard cross section calculated for a non-zero 
$k_T$~\cite{Cahn:1978se}.
The $\langle p_{T}^{h \,2}\rangle$ dependence on $z$ is built in and a 
suitable dependence of $\langle k_{T}^2\rangle$ 
on $x$ has been used to reproduce 
the values of $A^{UU}_{\cos \phi_h}$ at $z\lesssim 0.5$.
The agreement is satisfactory and the trends are similar over all 
bins, except for the two bins at  $p_T^h>0.5$ GeV/c and  $z>0.7$.

  The same Monte Carlo simulation 
is also  used to investigate the twist-4 
$\cos 2\phi_h$ azimuthal  modulations generated by the Cahn effect. 
The resulting amplitudes $A^{UU}_{\cos 2\phi_h}$ turn out to be compatible with 
zero. 
Other contributions, which are not generated by Boer-Mulders
and Collins effect, appear also at twist-4 or higher orders.
Although these contributions are not very well known, they should 
be suppressed as $1/Q^2$, thus it is most likely that the
 non-zero $A^{UU}_{\cos 2\phi_h}$ values of Fig.~\ref{fig:sub_cos2phi} 
are an indication of a non-zero Boer-Mulders PDF. 
Specifically, the corrected data for $A^{UU}_{\cos 2\phi_h}$ for positive hadrons
still show a strong $z$ dependence in the highest $p_T^h$-bin, 
with a significance above 5~$\sigma$. 
The phenomenological study of this effect is, however, beyond the scope of the
present paper.

\section{Conclusions}\label{sec:conclusions}
The COMPASS Collaboration has measured the azimuthal modulations of
positive and negative
hadrons from the decay of exclusive vector mesons produced in 
the scattering
of 160~GeV/c muons on a $^6$LiD target. The amplitudes of 
the modulations are found to be large
and of the same sign for positive and negative hadrons.
These hadrons constitute a contamination to the SIDIS hadron sample. 
Their contribution to the previously published COMPASS $A^{UU, \, publ}_{\cos \phi_h}$ and 
$A^{UU, \, publ}_{\cos 2\phi_h}$ unpolarised azimuthal asymmetries is estimated 
quantitatively and shown to be non-negligible over all the explored 
kinematic region and in particular at large $z$. 
After subtracting their $\cos \phi_h$  amplitudes, 
the $A^{UU}_{\cos\phi_h}$ asymmetries turn out 
to be in reasonable agreement over most of the explored kinematic region 
with a Monte Carlo simulation implementing the Cahn effect, except for a 
very few bins at large $z$ and large $p_T^h$.
The experimental determination of this important correction 
to already published data, which so far was never evaluated, 
is expected to have significant impact onto phenomenological analyses.
When implemented, it 
could hopefully allow for a successful disentangling of the various 
contributions to the data and for a first extraction of the 
Boer-Mulders function.\\

{\bf Acknowledgements}

We gratefully acknowledge the support of the CERN management and staff and
the skill and effort of the technicians of our collaborating institutes.
This work was made possible by the financial support of our funding
agencies.


\end{document}